\documentclass[10pt]{iopart}
\usepackage{graphicx}
\usepackage[latin1]{inputenc}
\usepackage[english]{babel}
\expandafter\let\csname equation*\endcsname\relax
\expandafter\let\csname endequation*\endcsname\relax
\usepackage{bm,amsmath,amssymb,epsfig,color}
\usepackage{iopams}
\begin{document}
\paper[]{Equilibrium time-correlation functions of
the long-range interacting Fermi-Pasta-Ulam model}
\author{P Di Cintio$^{1,2}$, S Iubini$^{3,4}$, S Lepri $^{4,2,6}$ and R Livi$^{5,2,4,6}$}
\address{$^{1}$Consiglio Nazionale delle Ricerche, Istituto di Fisica Applicata ``Nello Carrara'' via Madonna del piano 10, I-50019 Sesto Fiorentino, Italy}
\address{$^{2}$Istituto Nazionale di Fisica Nucleare, Sezione di Firenze,  
via G. Sansone 1 I-50019, Sesto Fiorentino, Italy}
\address{$^{3}$ Dipartimento di Fisica e Astronomia, Universit\`a di Padova, via F. Marzolo 8 I-35131, Padova, Italy}
\address{$^{4}$ Consiglio Nazionale
delle Ricerche, Istituto dei Sistemi Complessi, 
via Madonna del Piano 10, I-50019 Sesto Fiorentino, Italy}
\address{$^{5}$ Dipartimento di Fisica e Astronomia, Universit\`a di Firenze, 
via G. Sansone 1 I-50019, Sesto Fiorentino, Italy}
\address{$^{6}$ Centro Interdipartimentale per lo Studio delle Dinamiche Complesse,
Universit\`a di Firenze.}
\ead{p.dicintio@ifac.cnr.it,stefano.iubini@unipd.it,\\
stefano.lepri@isc.cnr.it,roberto.livi@unifi.it}
\begin{abstract}
We present a numerical study of dynamical correlations (structure factors) 
of the long-range generalization of the Fermi-Pasta-Ulam oscillator chain, where 
the strength of the interaction between two lattice sites
decays as a power $\alpha$ of the inverse of their distance.
The structure factors at finite energy density display distinct peaks,
corresponding to long-wavelength propagating modes, whose dispersion relation is compatible
with the predictions of the linear theory. We demonstrate that dynamical scaling holds, 
with a dynamical exponent $z$ that depends weakly on $\alpha$ in the range 
$1<\alpha<3$. The lineshapes have a non-trivial 
functional form and appear somehow independent of $\alpha$. 
Within the accessible time and size ranges, we also 
find that the short-range limit is hardly attained even for relatively large
values of $\alpha$.   
\end{abstract}
\submitto{Journal of Physica A: Mathematical and Theoretical}
\section{Introduction}
Statistical mechanics of long-range interacting systems displays many peculiar 
features like ensembe inequivalence, long-living  metastable states and 
anomalous diffusion of energy \cite{Bouchet2010,RuffoRev,Campa2014}. Other
unusual effects range from lack of thermalization upon interaction with a single 
external bath \cite{deBuyl2013} to the presence, in isolated systems, of non-isothermal inhomogeneous
stationary states, where the density and the temperature are anticorrelated 
\cite{2015PhRvE..92b0101T,gupta2016surprises}. From the dynamical point 
of view, propagation of perturbations can occur with infinite velocities, in a way
qualitatively different from the short-range cases \cite{Torcini1997,2005NYASA1045...68P,Metivier2014}.\\
\indent Long-range forces should have yet unexplored effects on energy transport 
for open systems interacting with external reservoirs. This issue has
so far received little attention in the 
literature \cite{avila2015length,Olivares2016,Bagchi2017,bagchi2017energy,Iubini2018} 
with respect to the case of short-range nonlinear, low-dimensional systems.
For the latter there is currently a detailed understanding  
of anomalous transport properties \cite{LLP03,Basile08,DHARREV,Iubini2012,Lepri2016}, 
leading to the breakdown of the classical Fourier law. Anomalous heat diffusion 
amounts to say that random motion of the energy carriers is basically  
a L\`evy walk \cite{Zaburdaev2015}, a description that accounts for most of the phenomenology 
\cite{Cipriani05,Lepri2011,Dhar2013}.\\
\indent A considerable insight has been obtained by Nonlinear Fluctuating Hydrodynamics (NFH) , whereby long-wavelength fluctuations are described in terms of hydrodynamic modes \cite{Spohn2014}. In a system with three conserved quantities, like chains of coupled oscillators with momentum conservation, the linear theory would yield two propagating sound modes and one diffusing heat mode, all of the three diffusively broadened. 
Nonlinear terms can be added and treated 
within the mode-coupling approximation \cite{Delfini07b,VanBeijeren2012,Spohn2014}.
This predicts that, at long times, the sound mode correlations 
satisfy the Kardar-Parisi-Zhang (KPZ) scaling, while the heat mode correlations 
follow a L\'evy-walk scaling. Several positive numerical tests 
for several models of coupled anharmonic oscillators with three conserved quantities 
(e.g., the Fermi-Pasta-Ulam chain with periodic boundary conditions) have been reported in the recent literature 
\cite{Das2014a,DiCintio2015,Mendl2013,Cividini2017}.\\
\indent A relevant consequence of the above approaches is that models can be classified 
in dynamical universality classes, mostly determined
by the conserved
quantities and the coupling among their fluctuations \cite{Popkov2015}. 
This entails the idea of  
dynamical scaling of equilibrium 
correlation functions and of the corresponding dynamical scaling exponent $z$ (defined below).
Thus, it is interesting to investigate for possible universality classes also in the
long-range models and 
seek for deviations from the standard diffusive behavior.\\
\indent In this paper we investigate how the interaction
range  exponent determines the scaling properties of equilibrium time-dependent correlations.
In the absence of a theoretical background, numerical results can be of guidance 
for constructing a theory: here 
we report a series of simulations for the Fermi-Pasta-Ulam
model with long-range interaction \cite{2014EL....10840006C}, previously investigated in different variants in the context of relaxation \cite{Miloshevich2015}, excitation propagation \cite{2005NYASA1045...68P,Miloshevich2017} and heat transport \cite{Olivares16,bagchi2017energy,bagchi2017thermal,Iubini2018}.\\
\indent 
The paper is organized as follows. Section \ref{model} describes the details of the model, 
while structure factors and their scaling properties are reported
in Section \ref{strufact}. The main features associated to the propagation of energy perturbations are discussed in
Section \ref{propag} together with the dependence of the standard chaos indicator, the maximum Lyapunov exponent, on the
range exponent in Section \ref{lyap}. The main results of our study are summarized in Section \ref{conclu}.
\section{The Model}
\label{model}
We consider a one-dimensional lattice of $N$ particles with periodic boundary conditions, whose dynamics is governed by the long-range Hamiltonian
\begin{equation}
\label{eq:hamilt}
H = \sum_{i=1}^N \left[\frac{p_i^2}{2} + 
\frac{1}{N_0(\alpha,N)}  \sum_{j\neq i}^{N} \,c_{ij}(\alpha) V(q_i-q_j) \right]
\end{equation}
where $q_i(t)$ and $p_i(t)$ are canonically conjugated variables (i.e., the displacement with respect to the
equilibrium position at the $i$-th lattice site and its associated momentum, respectively) 
and the function $V$ specifies the interaction potential. The strength of the interaction is controlled by the coupling matrix $c_{ij}(\alpha)=(d_{ij})^{-\alpha}$, where the quantity $d_{ij}$ identifies the shortest distance between sites $i$ and $j$ on a
periodic lattice~\cite{gupta2012overdamped,Gupta2014}, i.e. 
\begin{equation}
\label{eq:d}
d_{ij}=\min\{|i-j|,N-|i-j|\}
\end{equation}
The real non-negative exponent $\alpha$ is the parameter that controls the interaction range,
while $N_0(\alpha,N)$ is given by the generalized Kac prescription, that 
insures the extensivity with $N$ of Hamiltonian (\ref{eq:hamilt}):
\begin{equation}\label{eq:kac}
N_0(\alpha,N) = \frac{2}{N} \sum_{i=1}^N\sum_{j\neq i}^N c_{ij}(\alpha).
\end{equation}
\begin{figure}[ht]
\begin{center}
\hfill
\includegraphics[width=0.95\textwidth,clip]{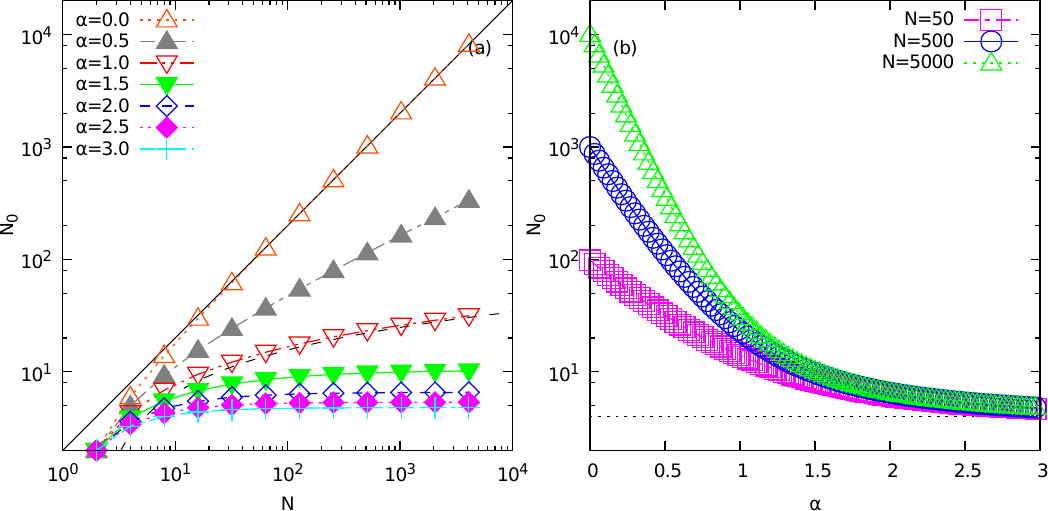}
\caption{(a) The Kac factor $N_0$ as function of the system size $N$. Different sets of points correspond to different values of the exponent $\alpha$ 
in the range $0\leq\alpha\leq 3$. The solid and the dashed lines draw the linear and the logarithmic trends for $\alpha=0$ and 1, respectively. 
(b) The Kac factor $N_0$ as function of the exponent $\alpha$ for $N=50$ (squares), 500 (circles) and 5000 (triangles). 
The thin dotted line draws the asymptotic value $N_0\to 4$ in the limit $\alpha\to\infty$.
}
\label{fign0}
\end{center}
\end{figure}
Notice that for $\alpha = 0$, i.e. the case of a fully connected lattice, one retrieves the standard Kac prescription, $N_0(0)=2(N-1)$. 
For any finite $\alpha$, $N_0(\alpha,N)$ is a monotonically increasing function of $N$: 
it has a finite positive derivative w.r.t. $N$
for  $0 < \alpha < 1$ ( i.e. in the region where Hamiltonian (\ref{eq:hamilt}) would be non-extensive in the absence
of the Kac factor), while in the large $N$ limit it converges to a constant for $\alpha > 1$.
The case $\alpha=1$ identifies the extensivity threshold in $d=1$ and $N_0(1,N)$ 
is characterized by a logarithmic divergence with $N$.
Finally, in the limit of $\alpha\to+\infty$ one obtains $N_0=4$ and $c_{ij}$ vanishes for $|i-j|>1$, while $c_{ij}= 1$ for $|i-j|=1$, thus retrieving the case of  nearest-neighbor interactions.
All this information is summarized in the two panels
of Fig. \ref{fign0}.

In this paper, we focus on the Fermi-Pasta-Ulam-$\beta$ (FPU) potential
\begin{equation}\label{fpupot}
V(x)=\frac{x^2}{2}+\frac{x^4}{4},
\end{equation}
for which we have assumed fixed dimensionless units, in such a way that the only relevant parameter is the total energy $H=E$, or,
equivalently, the energy per particle $e=E/N$. 
A few results about the model with the addition to $V$ of the cubic term $g|x|^3/3$ will be also reported. For $\alpha\to+\infty$ the above model reduces to the standard short-range FPU lattice, that has been
extensively studied in the context of heat transport in low-dimensional 
anharmonic chains \cite{LLP97,Lepri05,Wang2011}.

The linear dispersion relation for model Hamiltonian (\ref{eq:hamilt}) with $V$ given by (\ref{fpupot})
is obtained by neglecting the quartic term 
and looking for plane-wave solutions of the form $q_n\sim\exp(\imath kn-\imath \Omega_\alpha t)$ \cite{Miloshevich2015,Chendjou2018} :
\begin{equation}
\Omega_\alpha^2(k) = \frac{2}{N_0(\alpha)} \sum_{n=1}^N \frac{1-\cos kn}{n^\alpha}
\label{lindisp}
\end{equation}
In what follows we will consider periodic boundary conditions so that 
the allowed values of the wave number $k$ are integer multiples of $2\pi/N$.
Note that, at variance with \cite{Miloshevich2015,Chendjou2018}, the generalized Kac 
factor appears explicitly in the definition of $\Omega_\alpha$. An important feature of the linear dispersion relation is that in the small
wavenumber limit, $|k|\to 0$, the contribution of the  leading term is given by the proportionality relations  
\begin{equation}\label{lindisp2}
\Omega_\alpha(k) \propto |k|^{\frac{\alpha-1}{2}}, \quad \text{for} \quad 1<\alpha<3; \quad \propto |k|,  \quad \text{for} \quad \alpha\ge 3.
\end{equation}
As a consequence, the group velocity diverges as $|k|^{\frac{\alpha-3}{2}}$
in the first case, while it is finite in the second one. This result can be 
derived from the continuum limit of the equations of motion, where the 
long-range harmonic force can be approximated as a fractional 
derivative of order $(\alpha-1)$ \cite{Tarasov2006}.\\
\indent In the present work we will limit the analysis to the case $\alpha>1$
which is the most relevant to the aim of understanding the effect of 
the interaction range on heat-transport. 
In fact, in a previous paper \cite{Iubini2018} we collected evidence that 
in the genuine long-range case, $0 < \alpha<1$,
the mechanism of heat transport is dominated by the interaction of each oscillator with 
the external reservoirs, while the energy exchanged between oscillators is practically immaterial
in the limit of large values of $N$.
Conversely,  for $\alpha>1$
energy currents need to flow through the whole chain and bulk transport processes become relevant. It is thus important, to assess the type of energy diffusion that occurs there.\\
\indent Before discussing the main results, we want to comment about the numerical method 
we have adopted.
The forces acting between oscillators have been computed by an algorithm based on the Fast Fourier Transform, akin to the 
one previously used for similar models \cite{Gupta2014}. In fact, the form of 
the long-range potential defined  at the beginning of this Section allows to write forces as 
convolution products. This provides  a considerable advantage: the computational
cost to compute forces in a chain of $N$ oscillators amounts to ${\mathcal O}(N \ln N)$, to be compared  with any naive algorithmic implementation, that would demand  ${\mathcal O}(N^2)$ operations.
The integration of the equations of motion 
$\dot{p}_i = -\partial H / \partial q_i$ and $\dot{q}_i = \partial H / \partial p_i$ has been performed by
a 4-th order symplectic algorithm \cite{mclachlan1992accuracy} with fixed time step $\delta t =0.01$, 
that guarantees energy conservation with a relative accuracy of ${\mathcal O}(10^{-5})$.   
\section{Structure factors}
\label{strufact}
\begin{figure}[ht]
\begin{center}
\hfill
\includegraphics[width=0.8\textwidth,clip]{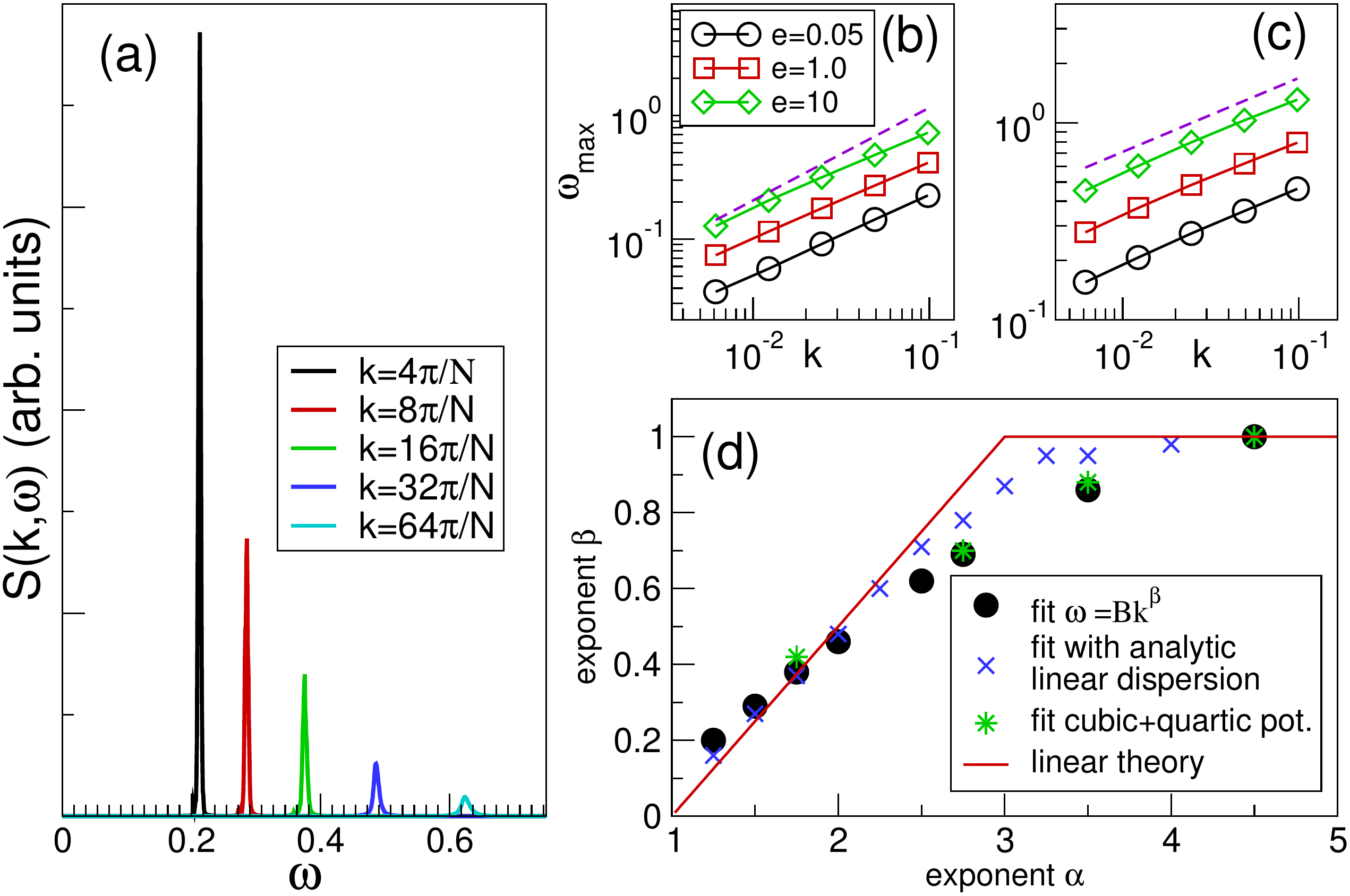}
\caption{(a) Structure factors of the displacement variable defined in Eq.(\ref{somega})
for the quartic potential, $N=4096$, $\alpha=1.75$, energy density $e=1$ and 
different values of the wavenumber $k$. Each peak is obtained by averaging over
$10^3$ independent dynamical trajectories lasting over $10^4$ time units.
(b,c) Dispersion relations obtained by plotting the peak frequency 
$\omega_{\rm max}(k)$ as a function of 
the wavenumber $k$, for $\alpha=1.25$ (b) $\alpha=1.75$ (c) 
and for different values of the energy
density $e$. The dashed lines 
are the power-law $|k|^{\frac{\alpha-1}{2}}$ predicted by the linear theory (\ref{lindisp2}). 
(d) The measured exponent $\beta$ versus $\alpha$, see text for details.
The solid line corresponds to the scaling given by (\ref{lindisp2}).
}
\label{fig:disper}
\end{center}
\end{figure}

In the spirit of the NFH theory, many interesting aspects of the heat transport mechanisms can be 
investigated by looking at the dynamical scaling of the structure factors associated to different
linear modes. To accomplish this task we consider the discrete space-Fourier transform of 
the particles displacement,
\begin{equation}
\hat q({k},t)=\frac{1}{N}\sum_{l=1}^N q_l(t)\,\exp(-\imath {k}l)
\end{equation}
and define the dynamical structure factor $S(k,\omega)$ as the ensemble-averaged 
modulus squared of the temporal Fourier transform 
of $\hat q({k},t)$:
\begin{equation}\label{somega}
S({k},\omega)=\langle|\hat q({k},\omega)|^2\rangle.
\end{equation}
Here the angular brackets denote an equilibrium average in the microcanonical ensemble characterized by the energy density $e$. 
It is worth recalling that, according to the Wiener-Khinchin theorem, $S({k},\omega)$
is the Fourier transform of the temporal autocorrelation function of $\hat q({k},t)$.
In the numerical implementation, the microcanonical equilibrium average can be estimated by evolving the dynamics over a sufficiently
large set of independent trajectories. A representative example of the numerical results is shown in Fig.\ref{fig:disper}. 
In the large-scale limit, i.e. for $k\to 0$, $S(k,\omega)$ 
exhibits sharp peaks at $\omega = \pm\omega_{\rm max}(k)$ (in panel (a) of Fig.\ref{fig:disper}
we display only the positive $\omega$-axis),
that correspond to some kind of propagating modes 
akin to sound modes usually observed in 
short-range interacting oscillators \cite{Lepri98c,Lepri2005,Gershgorin2005,Kulkarni2015}.  Note that there are no components around $\omega=0$. In the language of NFH 
this can be an indication that the heat-like mode does not couple significantly 
with the sound mode.\\
\indent In order to characterize the nature of these propagating excitations, in Fig.\ref{fig:disper}  (b), (c) and (d) 
we report an analysis of the positions of the peaks in $k$-space.
The figures in panels (b) and (c) show that  indeed $\omega_{\rm max}(k) $ scales with a power law $|k|^\beta$,
with $\beta$ roughly independent of the energy density. In this way one can estimate the value of $\beta$
and report it as a function of the exponent $\alpha$. The full circles in  Fig.\ref{fig:disper} (d) have been obtained
in this way. We can see that they are quite close to the predictions of the linear theory (Eq.(\ref{lindisp2}) ),
although some sensible deviations are present in the range $2 < \alpha < 4$.  Numerical data
can be better fitted by the function $\omega_{\rm max}(k) = B \Omega_\beta(k)$ with $\Omega_\beta$ 
given by Eq.(\ref{lindisp}): $B$ and $\beta$ are the fitting parameters. The estimates of $\beta$ obtained
in this way correspond to the crosses in  Fig.\ref{fig:disper} (d), that are closer to the linear scaling (\ref{lindisp2}).\\
\indent
Another interesting observation is that by adding to the interaction potential (\ref{fpupot})
the cubic term $g|x|^3/3$ the dependence of $\beta$ on $\alpha$ does not change
significantly (see the stars in Figure \ref{fig:disper} (d) ). We want to stress that the data
reported have been obtained for $e \sim \mathcal{O}(1)$, where the nonlinear terms of the
potential are by no means small with respect to the linear one.
This notwithstanding, we obtain evidence that the reasonable agreement of numerical
data with the linear dispersion relation accounts, upon 
a suitable energy-dependent parameter renormalization (by the  above defined constant $B$), 
for a characteristic speed of the excitations, as in the ``effective phonon"
description used for short-range interacting anharmonic lattices 
\cite{Lepri98c}.\\
\begin{figure}[ht]
\begin{center}
\hfill
\includegraphics[width=0.72\textwidth,clip]{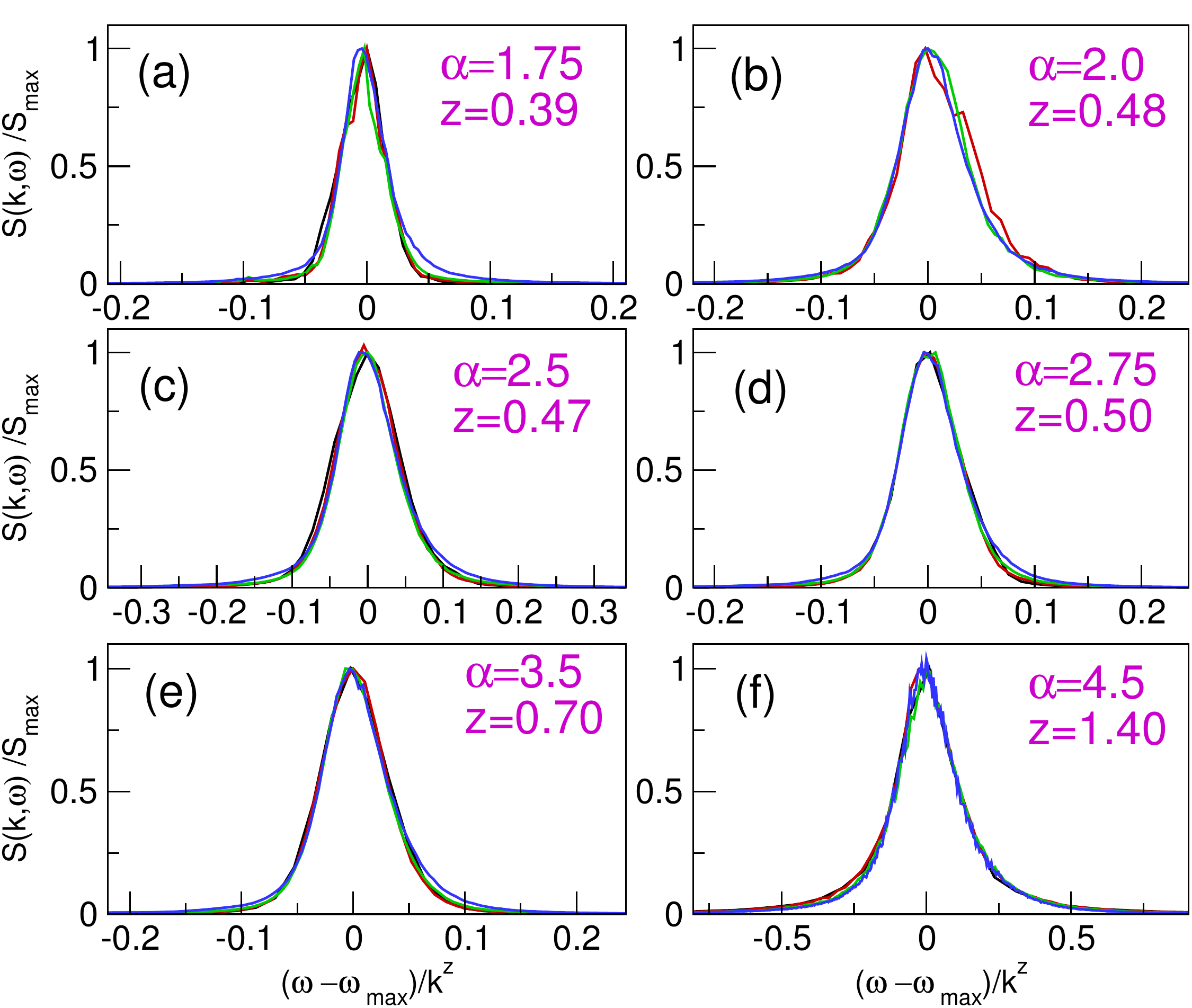}
\caption{Dynamical scaling of structure factors of displacement 
for the quartic model $N=2048$, energy density $e=1$ and 
different values of the range exponent $\alpha$. In each
panel four values of the wavenumber ($k=2\pi/N,4\pi/N,16\pi/N,32\pi/N$) 
are empirically collapsed according to formula (\ref{scaling32}), the 
best estimate of the dynamical exponent $z$ is reported alongside.
Data are averaged over  at least $10^3$ trajectories in the 
microcanonical ensemble.}
\label{fig:scale}
\end{center}
\end{figure}

\begin{figure}[ht]
\begin{center}
\hfill
\includegraphics[width=0.75\textwidth,clip]{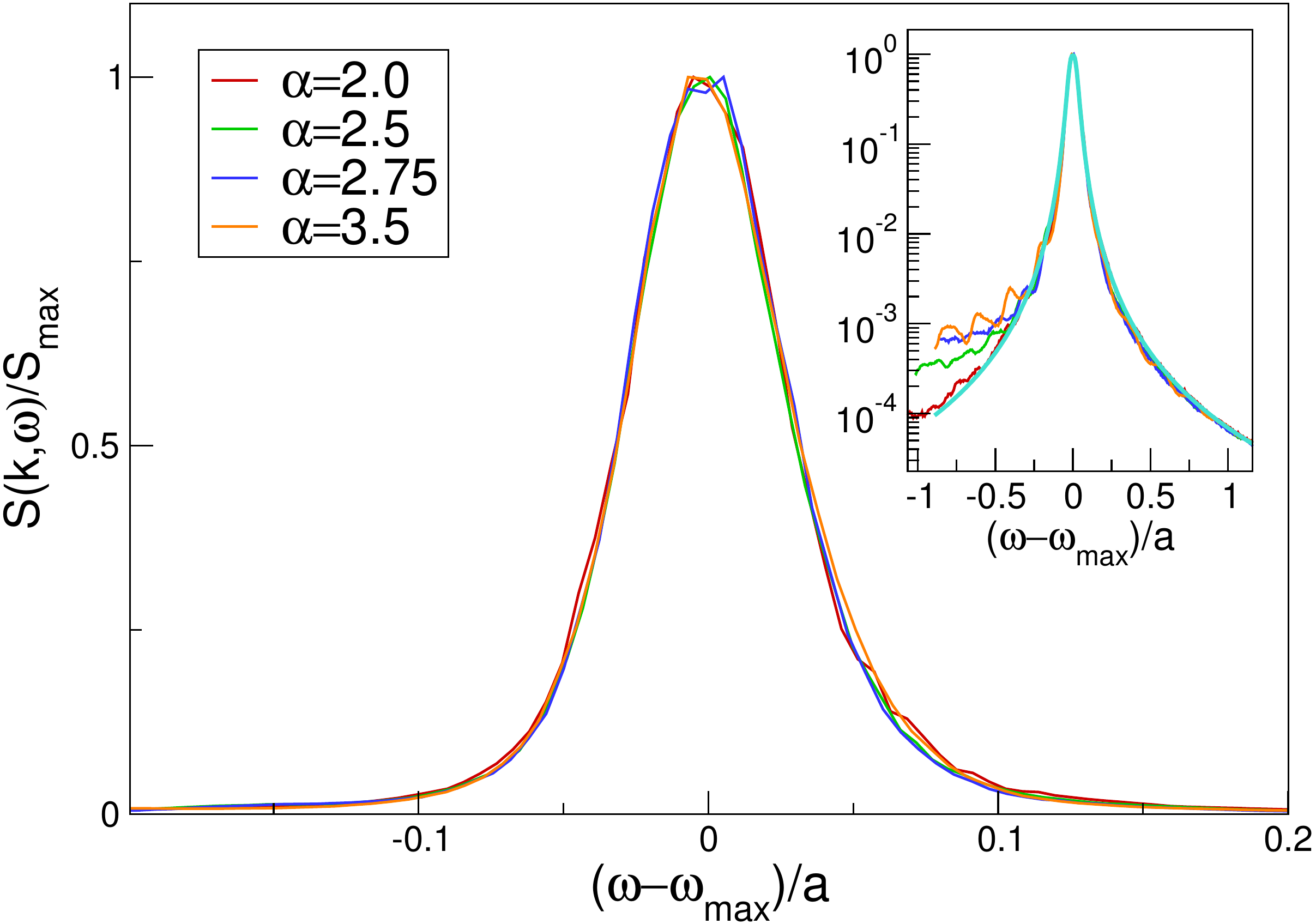}
\caption{Main panel: Line-shape of structure factors of displacement 
for the quartic potential; $N=2048$, energy density $e=1$ and fixed 
$k=16\pi/N$. A data collapse of line-shapes corresponding to different
values of the range exponent $\alpha$
is obtained by scaling the horizontal axis by a suitable factor $a$.
Inset: the same data plotted in semi-logarithmic scale:
the thick cyan solid line is a best fit with a function
$A/(B+x^\eta)$ with $\eta=2.77$. 
}
\label{fig:line}
\end{center}
\end{figure}

\begin{figure}[ht]
\begin{center}
\hfill
\includegraphics[width=0.72\textwidth,clip]{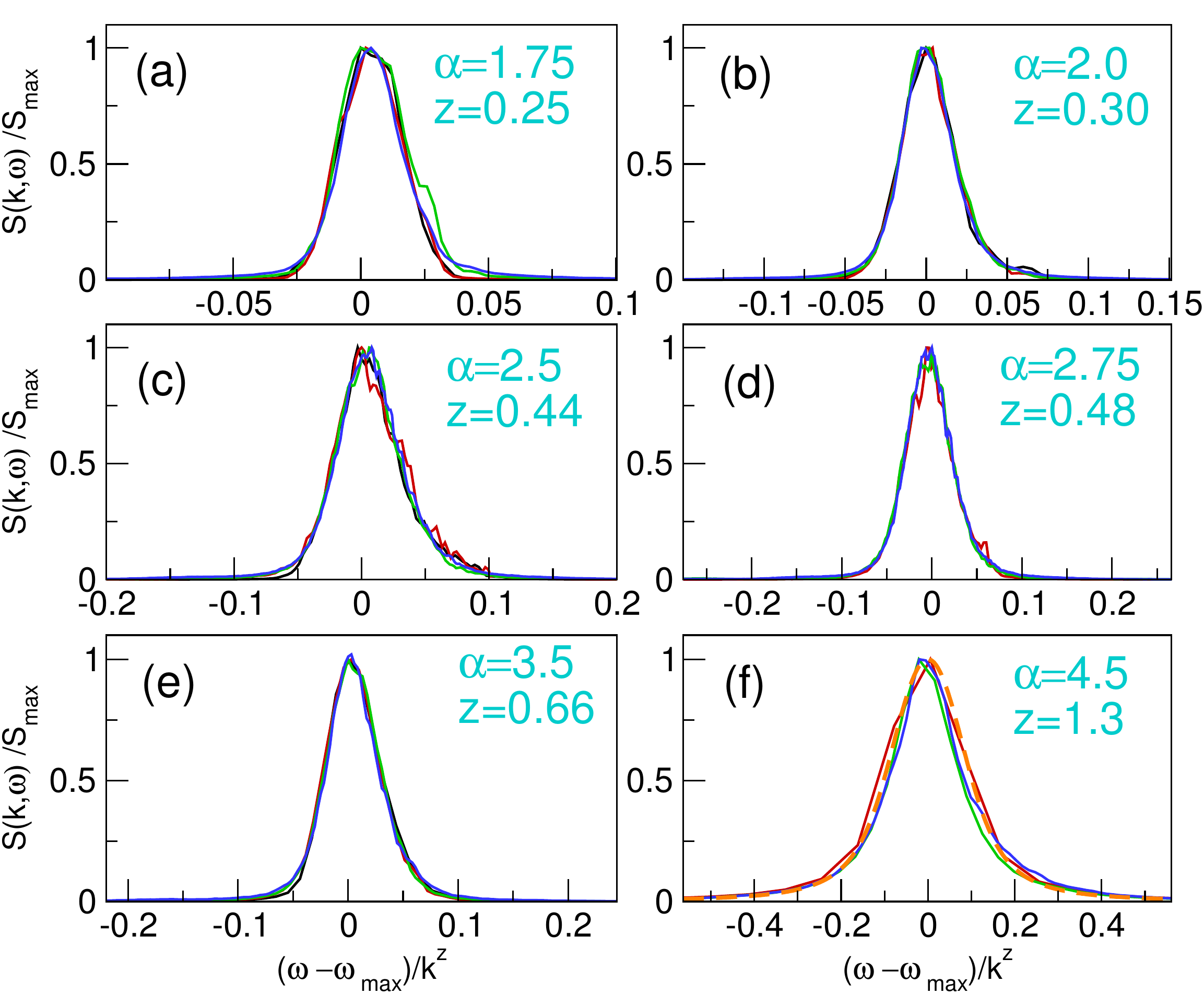}
\caption{Dynamical scaling of structure factors of displacement 
for the cubic plus quartic model $N=2048$, energy density $e=1$, $g=0.5$ and 
different values of the range exponent $\alpha$. In each
panel four values of the wavenumber ($k=2\pi/N,4\pi/N,16\pi/N,32\pi/N$) 
are empirically collapsed according to formula (\ref{scaling32}), the 
best estimate of the dynamical exponent $z$ is reported alongside.
Data are averaged over at least $10^3$ trajectories in the 
microcanonical ensemble. For comparison in panel (f) we plot 
the (suitably rescaled) function $f_{KPZ}$ predicted in 
the short-range case by NFH theory \cite{Spohn2014}
(dashed orange line).}	
\label{fig:scaleab}
\end{center}
\end{figure}

\indent Let us now turn to the issue of dynamical scaling.
In analogy with what found
in the short-range case, we may surmise that for $\omega\approx\pm \omega_{\rm max}$ 
structure factors of  
different wavenumbers $k$ are represented  by a suitable scaling function $f_\alpha$:
\begin{equation}
\label{scaling32}
S({k},\omega)\sim f_\alpha\left(\frac{\omega\pm\omega_{\rm max}}{k^{z}}\right).
\end{equation}
In this expression the subscript $\alpha$ points out that, in principle, the kind of scaling function 
$f$ might depend on $\alpha$. What is expected to depend on $\alpha$ is the dynamical exponent $z$.
This is a quantity of major importance, because its value determines the universality class of transport processes.
In Fig.\ref{fig:scale} we illustrate that the above surmise holds independently of 
$\alpha$. The data have been scaled empirically according to Eq.(\ref{scaling32})
and the best estimates of $z$ have been determined. The data collapse is generally
very good and in some cases excellent.\\ 
\indent A further important result is illustrated in Fig.\ref{fig:line}, where we compare the 
line-shapes of the structure factors for different values of $\alpha$ at fixed 
wavenumber $k$ (for the sake of clarity we report just the case $k=16\pi/N$). 
The line-shapes collapse very well onto each other by suitably rescaling the 
horizontal axis. Quite remarkably, this indicates that the form of the scaling function $f_\alpha$ should be
independent of  $\alpha$. In the absence of any theoretical hint on 
its functional form, in the inset of Fig.\ref{fig:line} we plot the same data in semi-logarithmic scale, along with an empirical fit. The available data rule out the possibility 
that the scaling function $f$ could be a simple standard lineshape, like a Gaussian or a Lorentzian one. Fitting rather suggests a non trivial behavior with slowly decaying tails. 
\\ 
\indent For comparison, we also performed a series of simulation for the FPU potential (\ref{fpupot})
with the addition of the cubic term $g|x|^3/3$. The results reported in Fig.\ref{fig:scaleab}
show that also in this case the scaling hypothesis works quite well
in the considered range of values of $\alpha$ and $k$. 
For this type of potential in the short-range case, the prediction of NFH is
$z=3/2$ and the scaling function $f_{\rm KPZ}$ 
is universal and known exactly \cite{Spohn2014}, albeit not in analytic form, so that one
has to compute it numerically \cite{Mendl2013}. In Fig.\ref{fig:scaleab} (f) 
($\alpha=4.5$) we plot for comparison also $f_{\rm KPZ}$:  it exhibits some
systematic deviations from the data-collapsed line-shape, while $z$  is still smaller than the one expected in the
short-range limit, $\alpha \to\infty$.\\
\begin{figure}[ht]
\begin{center}
\hfill
\includegraphics[width=0.8\textwidth,clip]{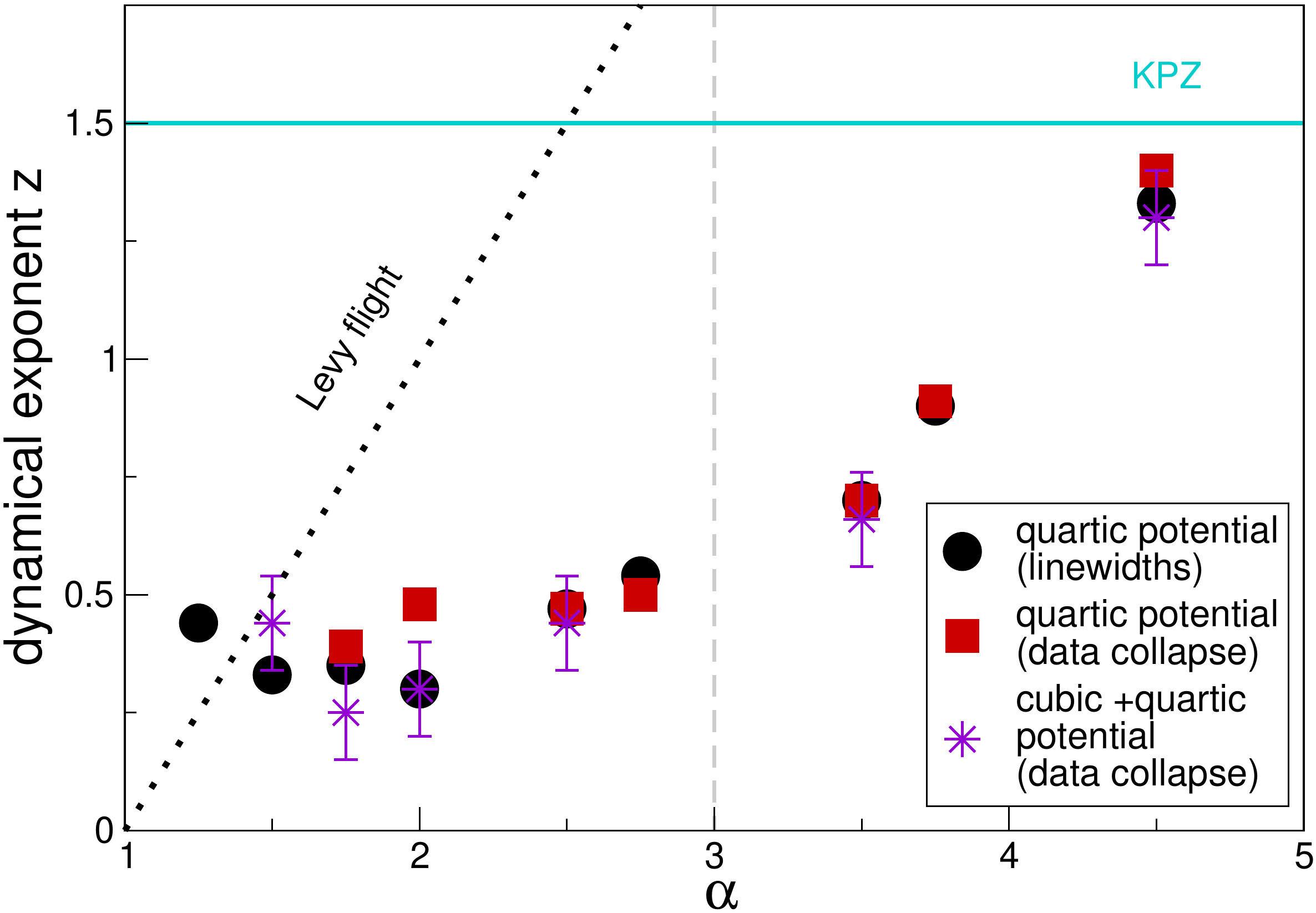}
\caption{Dynamical exponent $z$ extracted from the measurements of the sound peaks.
For comparison, the exponents are measured by data collapse (squares, stars) 
according to Eq.(\ref{scaling32}) and by fitting the dependence of line-widths 
on wavenumber at half maximum by a power law $k^z$ (circles). Stars refer to 
the FPU cubic plus quartic potential. The errors are tentative a priori estimates
of the empirical uncertainty in the data-collapse. The grey vertical line
signals the value $\alpha=3$ above which the group velocities of linear 
waves is finite.}
\label{fig:zeta}
\end{center}
\end{figure}
\indent The main outcome of our numerical analysis is that we have found  evidence that the dynamical exponent $z$ 
depends on the interaction range exponent $\alpha$ for both the FPU potential (\ref{fpupot})
and its cubic plus quartic variant.  The results are summarized in 
Fig.\ref{fig:zeta}, where, 
for comparison, we draw the function $z=\alpha-1$, which corresponds to the 
scaling relation for a simple L\'evy flight, i.e. a random walk with step length $\ell$ distributed 
with probability proportional to $\ell^{-\alpha}$ \cite{Bouchaud90}. The data show that the 
naive expectation that peak broadening might be described by such simple kinetic
process does not account for the observed dependence of $z$ on $\alpha$.\\
\indent Some further remarks are in order. First of all, for $1 < \alpha \lesssim 4$ the dynamical exponent $z$ is 
smaller than one. At first glance this could appear unusual and unexpected: for instance,  think about 
standard diffusion, where $z=2$. On the other hand, if one considers that 
the presence of long-range interactions induces instantaneous energy transfer, akin to the 
dynamical processes characterizing L\'evy flights, this fact seems less surprising. Moreover,
in the range $1<\alpha<3$, $z$ is weakly dependent on the range exponent $\alpha$,
and the numerical data could be also compatible with a constant value,  $z\approx 0.4$. 
Note also that the addition of the cubic term affects very little  the value of $z$, 
the differences being within the uncertainty of the empirical scaling procedure.
Finally, for $\alpha>3$ the exponents are remarkably smaller than 
what predicted for the short-range case within the NFH-mode-coupling approach \cite{Spohn2014},
i.e. $z=3/2$ (horizontal line in Fig.\ref{fig:zeta}) and $z=2$ for 
the cubic plus quartic and pure quartic potentials, respectively. 
In any case, data seem to indicate that the short-range limit
is approached very slowly and, accordingly,
we cannot exclude that such a case belongs to a different universality class.
\begin{figure}[ht]
\begin{center}
\hfill
\includegraphics[width=\textwidth,clip]{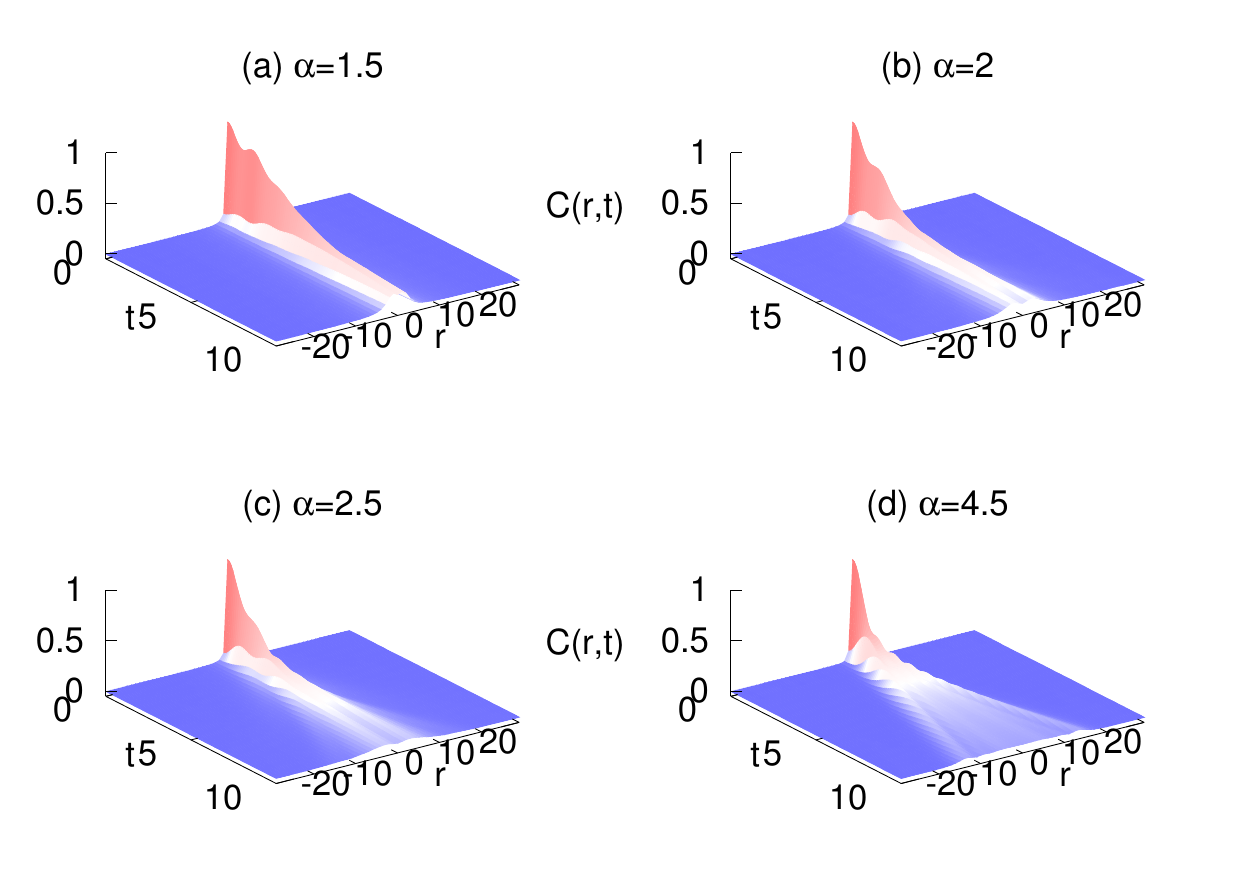}
\caption{The space-time excess energy correlations $C(r,t)$ for different values of $\alpha$.
Simulations have been performed  for a chain of $N=1024$ oscillators in a quartic potential with $e=1$.}
\label{fig:xcor}
\end{center}
\end{figure}
\section{Propagation of energy correlations}
\label{propag}
The structure factor of the displacement variables $q_i(t)$ gives direct information on 
propagating modes on long spatial and temporal scales. In heat 
transport problems one is also interested in the propagation of 
energy fluctuations. Further insight in the energy transport can be obtained by looking at the 
dynamics of the site energies 
\begin{equation}
h_i=\frac{p_i^2}{2} + 
\frac{1}{N_0(\alpha)}  \sum_{j\neq i}^{N} \,c_{ij} V(q_i-q_j)
\label{locen}
\end{equation}
and their (normalized) spatio-temporal correlation functions defined as 
\begin{equation}
C(r,t)=\frac{\langle h_{i+r}(t)h_i(0)\rangle -\langle h_i\rangle^2 }
{\langle h_{i}^2\rangle - \langle h_i\rangle^2}
\end{equation}
averaged over
the microcanonical ensemble (this is sometimes referred to as 
excess energy correlation \cite{Zhao06,Li2015}). As usual, translational invariance
is assumed, making $C(r,t)$ depend only on the relative distance $r$.\\
\indent In Fig.\ref{fig:xcor} we compare $C(r,t)$ for different values of $\alpha$.
The main outcome is that for  $\alpha<3$  energy spreading is somehow slower and
propagating peaks of excitation are lacking.
We also observed that a similar qualitative behavior characterizes the 
spreading of an initially localized finite  energy perturbation: for instance, this can
be checked for a perturbed thermal-equilibrium state, where the kinetic energies of the central 10 oscillators are perturbed (data not shown).
This indicates that for $1<\alpha < 3$, i.e. in the region of the parameter space where 
the linear group velocity diverges for small wavenumbers,
the model still retains some features of the pure long-range model, where energy can be trapped in single degrees of freedom for arbitrary long times
(e.g., see Ref. \cite{2005NYASA1045...68P}).
{We recall that a scaling analysis of the excess energy correlation for the a long-range model with harmonic nearest-neighbor coupling has been presented
in Ref.\cite{Bagchi2017}. Although this model has a different harmonic limit
(having finite group velocity) there are some resemblance with ours, including the presence of some propagating peaks.}
\begin{figure}[ht]
\begin{center}
\hfill
\includegraphics[width=0.7\textwidth,clip]{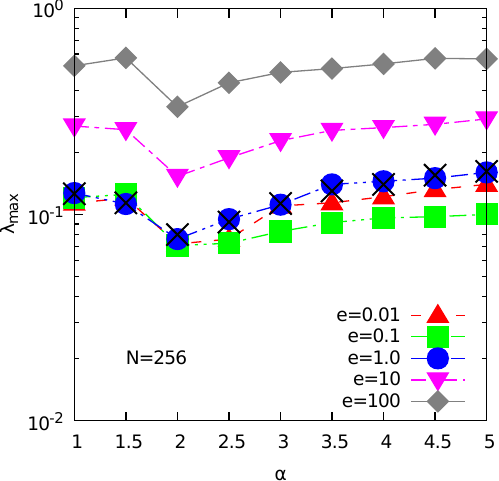}
\caption{Maximal Lyapunov exponent  $\lambda_{\rm max}$ for a long-range quartic FPU chain as function of the range exponent $\alpha$, for $N=256$ and different values of the energy density $e=0.01$ (upward triangles), 0.1 (squares), 1 (circles), 10 (downward triangles) and 100 (diamonds). The heavy crosses mark $\lambda_{\rm max}$ for the cubic plus quartic case with $e=1$.}
\label{figlyap}
\end{center}
\end{figure}

\indent We also performed some measurements of 
the energy structure factors $S_\mathcal{E}({k},\omega)$, that can be obtained by Fourier-transforming the energy density field 
defined in Eq.(\ref{locen}) and by computing the modulus squared of its temporal Fourier transform. 
Here we do not report numerical data, but we just comment  that  $S_\mathcal{E}({k},\omega)$ are 
characterized by a single central peak, whose width increases with the wavenumber $k$. This is qualitatively consistent 
with the spreading of the excess energy correlations. {Furthermore, 
a more quantitative analysis has been performed for a couple of values of $\alpha$: within the available frequency range, the data may compatible with dynamical scaling but a reliable estimate of dynamical exponents is not feasible. 
Also a tentative fitting, suggest that the $S_\mathcal{E}$ may have 
a Lorentzian lineshape, which would imply a L\'evy-function shape in real space, analogous to the short-range case \cite{Spohn2014}.
Surely a more} quantitative analysis 
would require very accurate statistical averages and we postpone this 
task to a future work.\\

\section{Lyapunov exponents}
\label{lyap}

\begin{figure}[ht]
\begin{center}
\hfill
\includegraphics[width=0.7\textwidth,clip]{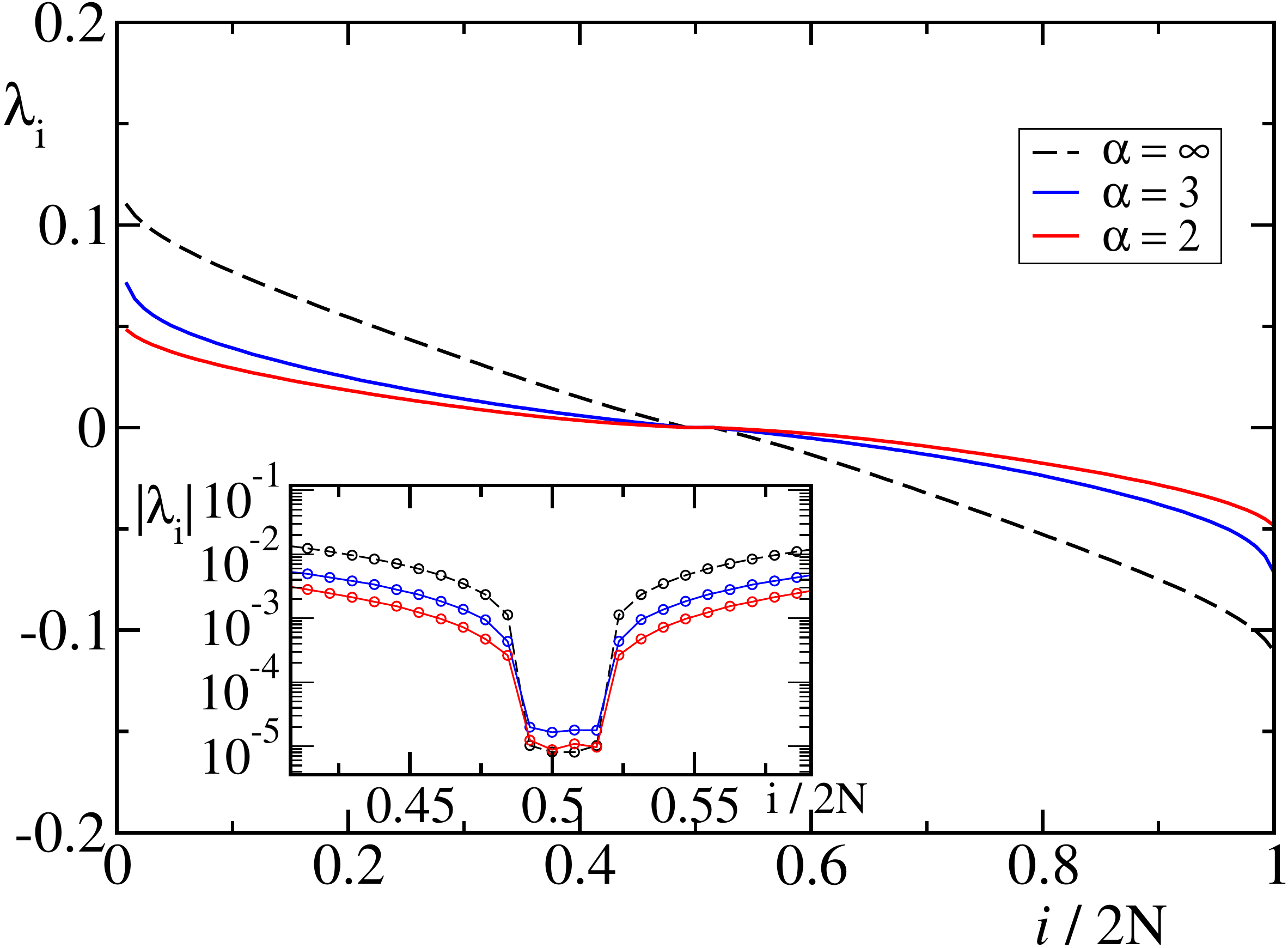}
\caption{Lyapunov spectra for the long-range quartic FPU chain $\alpha=2,3,\infty$, for $N=64$, energy density $e=1$. The inset reports the absolute values
of the exponents zoomed around its minima.}
\label{fig:lysp}
\end{center}
\end{figure}

\indent In this last section we complement the above results with an 
analysis of chaotic properties. Along with previous studies on similar long-range models (see e.g. Refs. \cite{christodoulidi2014,bagchi2017thermal,bagchi2017energy,2018EPJST.227..563C}) we have computed the maximal Lyapunov exponent $\lambda_{\rm max}$ for $1\leq\alpha\leq 5$ and $64\leq N\leq 16384$, making use of the standard Benettin-Galgani-Strelcyn technique \cite{Pikovsky2016}. If on one hand (as expected) $\lambda_{\rm max}$ is essentially constant for increasing $N$ at fixed energy density $e$, on the other hand, for fixed $N$ and for all values of the energy density explored, $\lambda_{\rm max}$ has a remarkably non-monotonic trend with $\alpha$. In particular, it has a relative minimum at $\alpha=2$, as shown in 
Fig. \ref{figlyap} for $N=256$ and different values of $e$ for the quartic case (filled symbols) and quartic plus cubic case (crosses). We note that, a similar behavior has been reported for the model in Ref. \cite{Bagchi2017}, where the quadratic term in 
(\ref{fpupot}) is outside the double sum in the model Hamiltonian (\ref{eq:hamilt}).  Interestingly, the $\alpha=2$ case stands out for exhibiting a 
seemingly ballistic behavior in heat transport \cite{Iubini2018}, that has led to speculate about some sort of (energy dependent) near-integrable behavior or the presence of additional conserved quantities. 
{To assess this possibility we computed also the entire Lyapunov spectra
$\lambda_i$, $i=1,\ldots,2N$
for $\alpha=2,3,\infty$. The data shown in Fig.~\ref{fig:lysp} demonstrate
that the case $\alpha=2$ has the usual four vanishing exponents as 
the others, corresponding to the 
usual conservation laws (see the inset of Fig.~\ref{fig:lysp}). 
So the existence of additional integrals of motion
should be ruled out. In agreement with this fact, both the scaling analysis of structure factors and the behavior of the space-time excess energy correlations do not display any particular feature to be singled out  for $\alpha=2$.}

\indent To conclude, let us also mention that the non monotonic behaviour of $\lambda_{\rm max}$ with a control parameter $\alpha$ at fixed energy density $e$ has been observed also for the short-range model obtained by adding to the Toda Hamiltonian a term proportional to $\sum_i |q_i|^\alpha$ (see \cite{2018arXiv180107153L,2018CSF...117..249D,2018arXiv181211770D}). In this case however, contrary to what observed here for the long-range FPU chain, such non monotonicity (again with a relative minimum for $\alpha=2$) disappears for increasing values of $e$ at fixed $N$ (see Fig. 2 in \cite{2018CSF...117..249D}), thus pointing towards a different origin of such a non-trivial behaviour of the degree of chaoticity in the two models.\\

\section{Conclusions}
\label{conclu}
In the present work we have undertaken a numerical study of some equilibrium 
correlations of the long-range FPU model with power-law decaying interaction strengths.
In particular, the structure factors of the displacement field provide an interesting complex 
scenario that we summarize hereafter.
\begin{itemize}
\item Even for values of the energy density corresponding to a strongly anharmonic regimes, 
we obtain convincing evidence of the existence of long-wavelength propagating modes,
whose dispersion relation is essentially the one valid for linear waves. 
\item We obtain also evidence of dynamical scaling, but the corresponding 
dynamical exponent $z$ depends on the interaction range exponent $\alpha$. In particular,
for $1< \alpha \lesssim 4$, $z$ is definitely smaller than one. Lacking any suitable theoretical
argument, we cannot envisage any simple relation between these two exponents. 
Even if it is reasonable to expect that $z$ eventually 
approaches its value in the short-range case (i.e. in the limit $\alpha\to\infty$), 
the convergence seems pretty slow.
\item Within the numerical accuracy of our simulations, we can conclude
that the line-widths of the structure factors are independent of $\alpha$, while
any standard Gaussian or Lorentzian form for the scaling function has to be ruled out. 
Moreover, the similarity between
the long-range models of the quartic and of the cubic plus quartic FPU potentials 
hints at some form of universality in the underlying effective non-linear
hydrodynamics. This is a bit surprising, if one considers that these two models,
in their short-range version,  belong to different universality classes, 
due to the different symmetries of the forces \cite{Lepri03,Spohn2014}.
Anyway, the previous conjecture demands to be checked for interaction potentials
other than the FPU ones -- a task that goes beyond the aims of this paper.

\item For what concerns the propagation of perturbations in these long-range models we have pointed out that there is a crossover from a localized regime to a propagating one when  $\alpha$ increases. A
more careful characterization of these two different dynamical phases certainly demands a 
further numerical effort.

\item The case $\alpha =2$ deserves some special consideration. 
The most puzzling aspect of this case is that different
versions of the long-range FPU quartic problem exhibit a sort of ``ballistic" transport for 
$\alpha =2$ (the same value, where $\lambda_{\rm max}$ exhibits a relative minimum).
The same peculiar feature does not show up for the cubic plus quartic case (data not shown), despite its overall similarity with the pure quartic one, even 
with respect to the non-monotonic trend of $\lambda_{\rm max}$ with $\alpha$. On the other hand, the dynamical exponent $z$
measured above is definitely different from one, the value one would expect for 
ballistic propagation. We do not have an explanation for such apparently 
contradictory behavior for equilibrium and non-equilibrium, that should 
be further explored in a future work.

\end{itemize}

\indent 

\section*{Acknowledgements}
SL acknowledges A. Torcini for useful discussions
and hospitality at the \textit{Laboratoire de Physique Th\'eorique et Mod\'elisation - LPTM} 
Cergy-Pontoise University and the \textit{Institut d'\'etudes avanc\'ees - IEA}
where part of this work has been undertaken.
SI acknowledges support from Progetto di Ricerca Dipartimentale BIRD173122/17.
\section*{References}
\bibliography{heat,bibliojpa}
\bibliographystyle{iopart-num}
\end{document}